\documentclass[]{interact}

\usepackage{epstopdf}
\usepackage[caption=false]{subfig}

\usepackage{algorithm}
\usepackage{algorithmic}
\usepackage{url}

\usepackage[numbers,sort&compress]{natbib}
\usepackage{color}
\usepackage{mathrsfs}
\bibpunct[, ]{[}{]}{,}{n}{,}{,}

\theoremstyle{plain}

\theoremstyle{definition}

\theoremstyle{remark}

\begin{document}

\articletype{article}

\title{Monitoring SEIRD model parameters using MEWMA for the COVID-19 pandemic with application to the State of Qatar.}

\author{
\name{E.~L. Boone\textsuperscript{a}\thanks{CONTACT E.~L. Boone. Email: elboone@vcu.edu}, Abdel-Salam G. Abdel-Salam\textsuperscript{b}, Indranil Sahoo\textsuperscript{a}, Ryad Ghanam\textsuperscript{c}, Xi Chen\textsuperscript{d},
 and Aiman Hanif\textsuperscript{a}}
\affil{\textsuperscript{a}Department of Statistical Sciences and Operations Research, Virginia Commonwealth University, Richmond, Virginia, USA; \textsuperscript{b}Department of Mathematics, Statistics and Physics, College of Arts and Sciences, Qatar University, Doha, Qatar. abdo@qu.edu.qa; \textsuperscript{c}Department of Liberal Arts and Science, Virginia Commonwealth University in Qatar, Doha, Qatar.raghanam@vcu.edu; \textsuperscript{d}Grado Department of Industrial and Systems Engineering, Virginia Tech, Blacksburg, Virginia, USA.
}
}
\maketitle

\begin{abstract}
During the current COVID-19 pandemic decision makers are tasked with implementing and evaluating strategies for both treatment and disease prevention.  In order to make effective decisions they need to simultaneously monitor various attributes of the pandemic such as transmission rate and infection rate for disease prevention, recovery rate which indicates treatment effectiveness as well as the mortality rate and others.  This work presents a technique for monitoring the pandemic by employing an Susceptible, Exposed, Infected, Recovered Death model regularly estimated by an augmented particle Markov chain Monte Carlo scheme in which the posterior distribution samples are monitored via Multivariate Exponentially Weighted Average process monitoring.  This is illustrated on the COVID-19 data for the State of Qatar.
\end{abstract}


%
%
\section{Introduction}\label{sec:Intro}
%

Coronavirus Disease 2019 (COVID-19) \cite{Wu, Rezabakhsh} is a severe pandemic affecting the whole world with a fast spreading regime, requiring to perform strict precautions to keep it under control. As there are limited cure and target treatment at the moment, establishing those precautions become inevitable.  These limitations \cite{Giuliani} can be listed as social distancing, closure of businesses / schools and travel prohibitions \cite{Chinazzi}. 

Corona Virus is a new human Betacoronavirus that uses densely glycosylated spike protein to penetrate host cells. The COVID-19 belongs to the same family classification with Nidovirales, viruses that use a nested set of mRNAs to replicate and it further falls under the subfamily of alpha, beta, gamma and delta Co-Vis. The virus that causes COVID-19 belongs to the Betacoronavirus 2B lineage and has a close relationship with SARS species. It is a novel virus since the monoclonal antibodies do not exhibit a high degree of binding to SARS-CoV-2. Replication of the viral RNA occurs when RNA polymerase binds and re-attaches to multiple locations \cite{McIntosh, Fisher}.

Cases of COVID-19 started in December 2019 when a strange condition was reported among a cluster of patients in Wuhan, China. Within a few weeks of this, the COVID-19 virus had spread to different parts of the world. On January 20, 2020, the first case of COVID-19 was recorded in the United States; Italy reported its first confirmed case on January 31,
2020. With COVID-19 cases rising across the world, the governments were soon seen intervening in financial and healthcare sectors. In late January, 2020, the first U.S. travel restrictions were imposed on travel from China. Weeks later, additional travel bans were imposed on countries in Europe and the United Kingdom. The World Health Organization (WHO) declared COVID-19 a pandemic on March 11, 2020, with a total of more than 100,000 cases globally. As of January 26, 2021, the worldwide total number of confirmed COVID-19 cases eclipses 100 million with over 2.15 million deaths. This virus has a global mortality rate of 3.4\%, which makes it more severe than the flu. The elderly who have other pre-existing illnesses are succumbing more to the COVID-19. People with only mild symptoms recover within 3 to 7 days, while those with conditions such as pneumonia or severe diseases take weeks to recover. As of January 26, 2021, the recovery percentage of patients, for example, in China stands at 95\%. The global recovery percentage rate of COVID-19 is somewhere around 97\% \cite{World}.  

The main efforts in the literature is focused on model estimation and forecasting the dynamic nature of the COVID-19 pandemic versus monitoring the process. The Susceptible, Exposed, Infected, Recovered, Death (SEIRD) model is a common compartmental model used for modeling disease through a population and variants of it have been used in modeling the COVID-19 pandemic, such as \cite{Loli2020} for Italy, \cite{Bagal2020} for India and \cite{Ghanam2020} for the State of Qatar. For more on disease modeling in general, see \cite{May1991, Clancy2008, Jewell2009, Vynnycky2010}.

Other modeling approaches include a time-series model to analyze the outbreak of the pandemic \cite{Deb2020}, a time-varying Bayesian semi-parametric model to look at short-term projections of the pandemic \cite{Roy2020}. \cite{Han2020} studies the dynamic pattern of COVID-19 deaths over time. The impact of government intervention such as a lockdown has been studied for China in \cite{Wang2020} and for India in \cite{Bagal2020, Basu2020}. Monitoring of the mathematical process has been done for Ukraine's COVID-19 outbreak \cite{Kyrychko2020}; however the parameter estimates are not generated directly from the data and no clear monitoring scheme is presented.  As the vaccine is beginning to be distributed to the public, monitoring the pandemic is more important to decision makers as they can determine how the pandemic is progressing as well as any shocks to the system that may be problematic. The monitoring approach needs to be able to react quickly to any large shifts in the system.

The goal of this work is to develop a method of monitoring a pandemic using a base mathematical model, such as SEIRD, that can be quickly updated as soon as new information comes in and can ``signal'' if there is a change in the parameters of the mathematical model.  The literature does not seem to provide any approaches that meet this goal.  The big challenge in this problem is updating the parameters in an automated way and converting those parameters to something that can be monitored.  Here a Bayesian approach is taken for the parameter estimation via a sampling algorithm that will allow for quick updating, avoid particle depletion and from which the samples can be monitored using a standard process control regime.

This work is organized in the following manner.  Section~\ref{sec:SEIRD} introduces the SEIRD model specific to the State of Qatar, the mean model, and the likelihood used. Section~\ref{sec.R0} introduces the Reproduction number and illustrates its deficiency in this case.  This is followed by the Markov chain Monte Carlo sampling algorithm with particle augmentation to update the parameters at each time step in Section~\ref{sec:Sample}.  The Multivariate Exponentially Weighted Moving Average (MEWMA) monitoring approach is presented in Section~\ref{sec:MEWMA}.  The method is illustrated on real data from the State of Qatar in Section~\ref{sec:Data}, which shows how the monitoring can be employed to identify critical changes in the pandemic.  Finally, Section~\ref{sec:Discuss} provides a discussion of the method and provides some suggestions for implementation as well as possible areas for improvement.

%
%
\section{SEIRD Model}\label{sec:SEIRD}
Let $t$ be a time index that is the number of days from the first recorded case of COVID-19 in the population of interest, $S(t)$ be the number of subjects Susceptible at time $t$, $E(t)$ be the number of subjects Exposed at time t, $I(t)$ be the number of Infected (symptomatic) subjects at time $t$, $R(t)$ be the cumulative number of Recovered subjects at time $t$ and $D(t)$ be the cumulative number of subject Deaths at time $t$.  This can be modeled with the following system of ordinary differential equations:

\begin{eqnarray}\label{eq:Sys1}
\frac{d S(t)}{d t} &=& -\alpha S(t)E(t)  \cr
\frac{d E(t)}{d t} &=& \alpha S(t)E(t) - \beta E(t) - \gamma E(t) \cr
\frac{d I(t)}{d t} &=& \beta E(t) - \gamma I(t) - \eta I(t)   \cr
\frac{d R(t)}{d t} &=& \gamma I(t) \cr
\frac{d D(t)}{d t} &=& \eta I(t),  
\end{eqnarray}

\noindent where the parameters are explained as follows: $\alpha$ is the transmission rate (per day $\times$ individual$^2$) from Susceptible to Exposed, the rate (per day) at which Exposed become Infected (symptomatic) is denoted by $\beta$, $\gamma$ is the rate (per day) at which Infected become recovered and the mortality rate (per day) for those Infected is denoted by $\eta$.  Notice that, this model formulation makes several key assumptions which are as follows.  Immigration, emigration, natural mortality and births are negligible over the time frame and hence are not in the model.  Once a person is in the Infected group, they are quarantined and hence they do not mix with the Susceptible population.  The Recovered and Deaths compartments are for those who first are infected.  Those who are exposed and are asymptomatic recover at the same rate $\gamma$ as those who become sick and recover.  The SEIRD model presented here is the same as the one presented in \cite{Ghanam2020} and matches the assumptions needed for the example provided in Section~\ref{sec:Data}, however, the estimation and monitoring method is not specific to this particular model. 

Due to the dynamic nature of how the pandemic has developed, assuming the system is in ``steady state'' is invalid as governments have intervened into the system in an effort to influence various parameters as well as medical treatment of the disease has changed across the time frame.  Hence the parameters are also functions of time denoted as $\alpha(t),~\beta(t),~\gamma(t)$ and $\eta(t)$. Specifically,
\begin{eqnarray}\label{eq:Sys2}
\frac{d \lambda_{S}(t)}{d t} &=& -\alpha(t) \lambda_{S}(t)\lambda_{E}(t)  \cr
\frac{d \lambda_{E}(t)}{d t} &=& \alpha(t)  \lambda_{S}(t)\lambda_{E}(t) - \beta(t) \lambda_{E}(t) - \gamma(t) \lambda_{E}(t) \cr
\frac{d \lambda_{I}(t)}{d t} &=& \beta(t) \lambda_{E}(t) - \gamma(t) \lambda_{I}(t) - \eta(t) \lambda_{I}  \cr
\frac{d \lambda_{R}(t)}{d t} &=& \gamma(t) \lambda_{I}(t) \cr
\frac{d \lambda_{D}(t)}{d t} &=& \eta(t) \lambda_{I}(t),
\end{eqnarray}
where  $\lambda_{s}(t)$, $\lambda_{E}(t)$, $\lambda_{s}(I)$, $\lambda_{R}(t)$ and $\lambda_{D}(t)$ denote the respective mean parameters.

At each time point $t$ the parameters must have a prior distribution.  For this work the prior distribution specification will be the same for all $t$, however, this is not necessary if one has information that needs to be included at a specific time.  

Since a Bayesian methodology is being employed the likelihood is specified to be:
\begin{eqnarray}
  I(t) &\sim& Poisson \left( \lambda_I(t) \right) \cr
  R(t) &\sim& Poisson \left( \lambda_R(t) \right) \cr
  D(t) &\sim& Poisson \left( \lambda_D(t) \right) .
\end{eqnarray}

Notice that $S(t)$ and $E(t)$ are not in the likelihood as they are {\em latent} states in that they are not directly observed.  The true likelihood for $\lbrace S(t), E(t), I(t), R(t), D(t) \rbrace$ should be Multinomial.  However, with two latent states, one of which is the largest state, the Multinomial approach is challenging to apply, thus this work uses adopts Poisson likelihood as an approximation.

\section{The Basic Reproduction Number $R_{0}$}\label{sec.R0}

The basic reproduction number, $R_0$, is defined as the expected number of secondary cases produced by a single infection in a completely susceptible population. It is dimensionless and can be calculated as a product of the transmissibility, the average rate of contact between susceptible and infected individuals, and the duration of the infectiousness \cite{Chowell2009}.
In model (1), the last two equations do not contribute to $R_0$ and so
\[   R_0 =  \frac{\alpha}{\beta + \gamma }. \]
\noindent Since our model is time varying, it follows that
\[ R_0(t) = \frac{\alpha(t) }{\beta(t) + \gamma(t) } \]
In this work the Reproduction number is not considered since it does not account for all the parameters in the model.  One of the goals of this work is to ensure the monitoring process accounts for all of the parameters.

%
%

\section{Sequential Sampling with Particle Augmentation}\label{sec:Sample}

From a monitoring perspective, we need to essentially estimate the parameters,  $\theta_t =\lbrace \alpha(t), \beta(t), \gamma(t), \eta(t) \rbrace$), at each time step dependent primarily on the latest data.  If at time $t$, we want to estimate the change in the parameters from $t-1$ to $t$,  we need an estimation algorithm that can update the parameters so that changes can be detected.  In a dynamic system where the parameters are changing due to interventions into the system such as vaccines or quarantines, the estimation algorithm needs to also take into account the previous changes that have taken place in the system.  Traditional approaches to modeling varying coefficient models from a Bayesian perspective have difficulties as the sampling methods required often suffer from {\em particle depletion} as the process proceeds through time. 
 
The proposed Algorithm~\ref{alg:SIR} is a variant of the Sampling Importance Resampling algorithm at each time step.
To ensure there are enough particles to work through the sampling process, the basic idea is to augment each accepted sample by some random perturbations to generate new particles to move through to the next step.

Let us fix the notation first. Denote $\mathbf{D}(k) = \lbrace S(k), E(k), I(k), R(k), D(k) \rbrace$
as the actual state vector of the system at time step $k$, where the first two components are latent, $\forall k\in \{0,1,\ldots,T\}$ and $T$ denotes the total number of time steps considered. Denote $g(\theta_0)$ and $g(\theta_k|\theta_{k-1})$, respectively, as the candidate distributions to sample $\theta$ from at time step $0$ and at time step $k$ ($\forall k=1,2,\ldots, T$). Let $\tilde{\theta}_{k}$ denote the set of accepted samples of $\theta$ at time $k$ that will be passed on to the next time step ($\forall k\in \{0,1,\ldots,T-1\}$).  Denote the set that contains all the accepted values of $\theta$ up to time $k$ by $\tilde{\Theta}(k)$.

We elaborate on the key steps of  Algorithm~\ref{alg:SIR} below. 
At time step $0$, we draw $n_c$ candidate samples of $\theta$ from 
$g(\theta_0)$, denoted by $\{\theta_{0,j}^{*}\}_{j=1}^{n_c}$. 
 We then evaluate the posterior distribution using the data at time 0 (i.e., $\mathscr{D}(0)$) and the candidate particles  to obtain the unnormalized weights $\{w_{j}^{*}, j=1,2,\ldots, n_c\}$ at time step $0$, which are subsequently normalized to  the  $\widehat{w}_{j}^{*}$. 
We then obtain $n_p$ posterior samples by selecting from $\{\theta_{0,j}^{*}, j=1,2, \ldots, n_c\}$ with the corresponding probabilities given by $\{\widehat{w}_{j}^{*}\}_{j=1}^{n_c}$; denote this set of $n_p$ samples by $\tilde{\theta}_{0}=\{\tilde{\theta}_{0,1}, \tilde{\theta}_{0,2}, \ldots, \tilde{\theta}_{0,n_p}\}$. This set of accepted values of $\theta$ will be passed on to time step 1. 
The sampling for time step $k$ is similar to that for time step $0$ but enhanced with sample augmentation. Specifically,
at time step $k$ ($\forall k\in \{1,2,\ldots,T\}$), using an appropriate candidate density $g(\theta_k|\tilde{\theta}_{k-1, \ell})$ which is conditioned on each accepted sample from time step $k-1$,  
we generate a batch of $n_b$ candidate samples in the neighborhood of $\tilde{\theta}_{k-1, \ell}$; denote the batch by $\{\theta_{k,\ell,j}^{*}\}_{j=1}^{n_b}$ for $\ell = 1,2,\ldots, n_p$.  
We then evaluate the posterior distribution using the data up to time $k$  (i.e., $\mathscr{D}(k)$), the accepted samples up to time step $k-1$ (i.e., $\tilde{\Theta}(k-1)$) and the candidate particles, to obtain the unnormalized weights $\{w_{\ell, j}^{*}, j=1,2,\ldots, n_b, \ell = 1,2,\ldots, n_p \}$ at time step $k$, which are subsequently normalized to  the  $\widehat{w}_{\ell, j}^{*}$. 
We then obtain $n_p$ posterior samples by selecting from $\{\theta_{k,\ell,j}^{*}, j=1,2, \ldots, n_b, \ell = 1,2,\ldots, n_p\}$ with the corresponding probabilities given by $\{\widehat{w}_{\ell,j}^{*}, j= 1,2,\ldots,n_b, \ell = 1,2,\ldots, n_p\}$; denote this set of $n_p$  accepted samples by $\tilde{\theta}_{k}=\{\tilde{\theta}_{k,1}, \tilde{\theta}_{k,2}, \ldots, \tilde{\theta}_{k,n_p}\}$.  Then by the end of time step $k$, we update the set of accepted values of $\theta$ to $\tilde{\Theta}(k)= \tilde{\Theta}(k-1) \cup \tilde{\theta}_{k}$.

\begin{algorithm}
\caption{Enhanced sampling importance resampling algorithm}
\label{alg:SIR}
\begin{algorithmic}[1]
\STATE Specify a prior distribution $p_0( \theta_0 )$, and define $\mathbf{D}(0)=   \lbrace S(0), E(0), I(0), R(0), D(0) \rbrace $. Let $\mathscr{D}(0) = \mathbf{D}(0).$
\STATE Draw $n_c$ candidate samples from a candidate distribution $g(\theta_0)$, i.e., $\theta_{0,j}^{*} \sim g(\theta_0)$, $j=1,2,\ldots,n_c$, then $\mathbf{D}(1)$ is observed.
\STATE Evaluate $w_{j}^{*} = p(\mathbf{D}(1) | \mathscr{D}(0), \theta_{0,j}^{*} )p_0( \theta_{0,j}^{*})/g(\theta_{0,j}^{*})$, $j=1,2,\ldots,n_c$.
\STATE Obtain normalized weights $\widehat{w}_j^{*} =  w_j^{*} /\left(\sum_{\ell=1}^{n_c} w_\ell^{*}\right)$, $j=1,2,\ldots,n_c$.
\STATE Resample from $\{\theta_{0,j}^{*}\}_{j=1}^{n_c}$ using the set of normalized weights, $\{\widehat{w}_j^{*}\}_{j=1}^{n_c}$, and obtain a set of $n_p$ samples, $\tilde{\theta}_{0}=\{\tilde{\theta}_{0,1}, \tilde{\theta}_{0,2}, \ldots, \tilde{\theta}_{0,n_p}\}$. Let $\tilde{\Theta}(0)= \tilde{\theta}_{0}$.\\
\FOR{$k=1,2,\ldots,T$}
\STATE Update the total dataset with new observations, $\mathscr{D}(k) =  \mathbf{D}(k) \cup \mathscr{D}(k-1)$, where   $\mathbf{D}(k) = \lbrace S(k), E(k), I(k), R(k), D(k) \rbrace$.

\STATE Specify a prior distribution $p_k( \theta_k )$, then $\mathbf{D}(k+1)$ is observed.
\STATE For the $\ell$th sample in  the  collection $\tilde{\theta}_{k-1}$ from step $k-1$, $\tilde{\theta}_{k-1,\ell}$, draw $n_b$ samples from the candidate distribution $g(\theta_k | \tilde{\theta}_{k-1,\ell} )$, i.e., $\theta_{k,\ell,j}^{*} \sim g(\theta_k | \tilde{\theta}_{k-1,\ell} )$, $j=1,2,\ldots, n_b$.

\STATE Evaluate   
$w_{\ell,j}^{*} = p( \mathbf{D}(k+1) | \mathscr{D}(k), \theta_{k,\ell,j}^{*}, \tilde{\Theta}(k-1))p_k(\theta_{k,\ell,j}^{*})/g( \theta_{k,\ell,j}^{*} | \tilde{\theta}_{k-1,\ell} )$, $\ell = 1,2,\ldots, n_p$, $j=1,2,\ldots, n_b$.

\STATE Normalize $\widehat{w}_{\ell,j}^{*} =   w_{j,\ell}^{*} / \left(\sum_{\ell=1}^{n_p}\sum_{h=1}^{n_b}  w_{h,\ell}^{*}\right)$, $\ell = 1,2,\ldots, n_p$, $j=1,2,\ldots, n_b$.

\STATE Resample from $\{\theta_{k,\ell,j}^{*},\ell =1, 2,\ldots, n_p; j=1, 2, \ldots, n_b\}$ using the set of normalized weights, $\{\widehat{w}_{\ell,j}^{*}, \ell =1, 2,\ldots, n_p; j=1, 2, \ldots, n_b\}$, and obtain a set of $n_p$ samples, $\tilde{\theta}_{k}=\{\tilde{\theta}_{k,1}, \tilde{\theta}_{k,2}, \ldots, \tilde{\theta}_{k,n_p}\}$. Let $\tilde{\Theta}(k)= \tilde{\Theta}(k-1) \cup \tilde{\theta}_{k}$.
 \ENDFOR
\end{algorithmic}
\end{algorithm}

%
%
\section{Monitoring}\label{sec:MEWMA}
The presented method for estimating the SEIRD model parameters through time is quite responsive to changes in the system.  Hence, these parameters can be monitored to look for changes in the parameters which will be manifested in the data.  Since there are four parameters in the SEIRD model that need to be simultaneously monitored, the Multivariate Exponentially Weighted Moving Average (MEWMA) approach was chosen as an appropriate monitoring method. \cite{Lowry} developed a multivariate EWMA (MEWMA) control chart, which is an extension to the univariate EWMA.  

First the parameter samples were differenced using a single backward lag of one, $\Delta \alpha(t)_i = \alpha(t)_i - \alpha(t-1)_i$, $\Delta \gamma(t)_i = \gamma(t)_i - \gamma(t-1)_i$, $\Delta \beta(t)_i = \beta(t)_i - \beta(t-1)_i$ and $\Delta \eta(t)_i = \eta(t)_i - \eta(t-1)_i$.  These form a vector $\left( \Delta \alpha(t)_i,~\Delta \gamma(t)_i,~\Delta \beta(t)_i,~\Delta \eta(t)_i \right)^{T}$ to be monitored for significant deviations from zero, which would correspond to a significant change in the set of parameters.  The multivariate parameters are given by the mean:
\[ \Delta \bar{\boldsymbol\theta}(t) = \frac{1}{n_p} \mathbf{1}_{n_p}^T \left( \Delta\boldsymbol\alpha(t), \Delta\boldsymbol\gamma(t), \Delta\boldsymbol\beta(t), \Delta\boldsymbol\eta(t) \right) \]
and variance:
\[ Cov( \Delta\boldsymbol\theta (t)) = \frac{1}{n_p-1}
\begin{pmatrix}
\Delta\boldsymbol\alpha(t)^T\\ \Delta\boldsymbol\gamma(t)^T\\ \Delta\boldsymbol\beta(t)^T\\ \Delta\boldsymbol\eta(t)^T
\end{pmatrix}
\left( \Delta\boldsymbol\alpha(t), \Delta\boldsymbol\gamma(t), \Delta\boldsymbol\beta(t), \Delta\boldsymbol\eta(t) \right).
\]
In order to have a monitoring process that is not too sensitive, the MEWMA is employed which is given by
\begin{equation}\label{eq:MWEMA1}
  MEWMA(t) = \Lambda \Delta \bar{\boldsymbol\theta}(t) + (1-\Lambda)MEWMA(t-1),
\end{equation}
with the moving covariance matrix:
\[ V(t) = \Lambda^2 Cov( \Delta\boldsymbol\theta (t)) + ( 1 - \Lambda )^2 V(t-1),  \]
where $\Lambda$ is a smoothing coefficient that controls how much a new observation can influence the overall mean.  Lower values of $\Lambda$ are more conservative in that new observations do not have much influence on the mean as higher values allow new observations to have a greater influence on the mean.  This is used to control false signals as any process being monitored will have some natural variation that may cause the observation to be signalled.  Typical values of $\Lambda$ include $0.1,~0.15,~0.2,~0.25$ and $0.3$.

Since we are looking at the differences in parameter values, the target value for the process mean should be $(0,0,0,0)$ indicating no shift in the process.  In the multivariate case, a test statistic based on Hotelling's $T^2$ can be formed as

\begin{equation}\label{eq:T2}
T^2(t) = MEWMA(t)^T V(t)^{-1} MEWMA(t).
\end{equation}
When $n_p$ is large, $T^2(t)$ in Equation (\ref{eq:T2}) is approximately $\chi^2(4)$ which has a 0.95-quantile equal to 9.48.  Hence any $T^2(t) > 9.48$ would be deemed as a significant change in the parameter differences and thus a significant change in the SEIRD process.  Note that, in our case $n_p = 1,000$ and hence deemed large enough for the $T^2$ approximation.

%
%
\section{Data Analysis: State of Qatar}\label{sec:Data}

\subsection{Data Description}

The World Health Organization, Johns Hopkins University and other agencies maintain data sets on the daily number of confirmed infected cases, deaths, and recoveries for every country. In this work, we study the evolution of the pandemic in the State of Qatar. All data for Qatar were obtained from Johns Hopkins University and are freely accessible via the Johns Hopkins COVID-19 GitHub repository \cite{Miller}. The GitHub site includes daily cumulative number of confirmed infections, cumulative number of recovered and cumulative number of deaths starting 22 January 2020. 

The goal of the data analysis is to demonstrate and assess the proposed modeling approach and its use in monitoring the pandemic as the varying coefficient approach will allow for the model parameters to adjust quickly to changes in the data generation process. In model (1), the Recovered and Death states are cumulative with no outgoing transitions whereas the Infected state has transitions from Exposed and to Recovered and Death states. Hence the data for confirmed infections are cumulative and include the numbers from both the Recovered and Death states. As such, if $CI(t)$ denotes the confirmed infections at time $t$, then the number of Infected subjects at time $t$ is defined as  
$$
I(t) = CI(t) - R(t) - D(t) .
$$
From here on, the term ``Active Infections'' will be used to denote this derived variable versus the cumulative Infected provided in the data.

\begin{figure}
\begin{center}
\begin{tabular}{ccc}
(a) & (b) & (c) \\
\includegraphics[width = 0.31\textwidth]{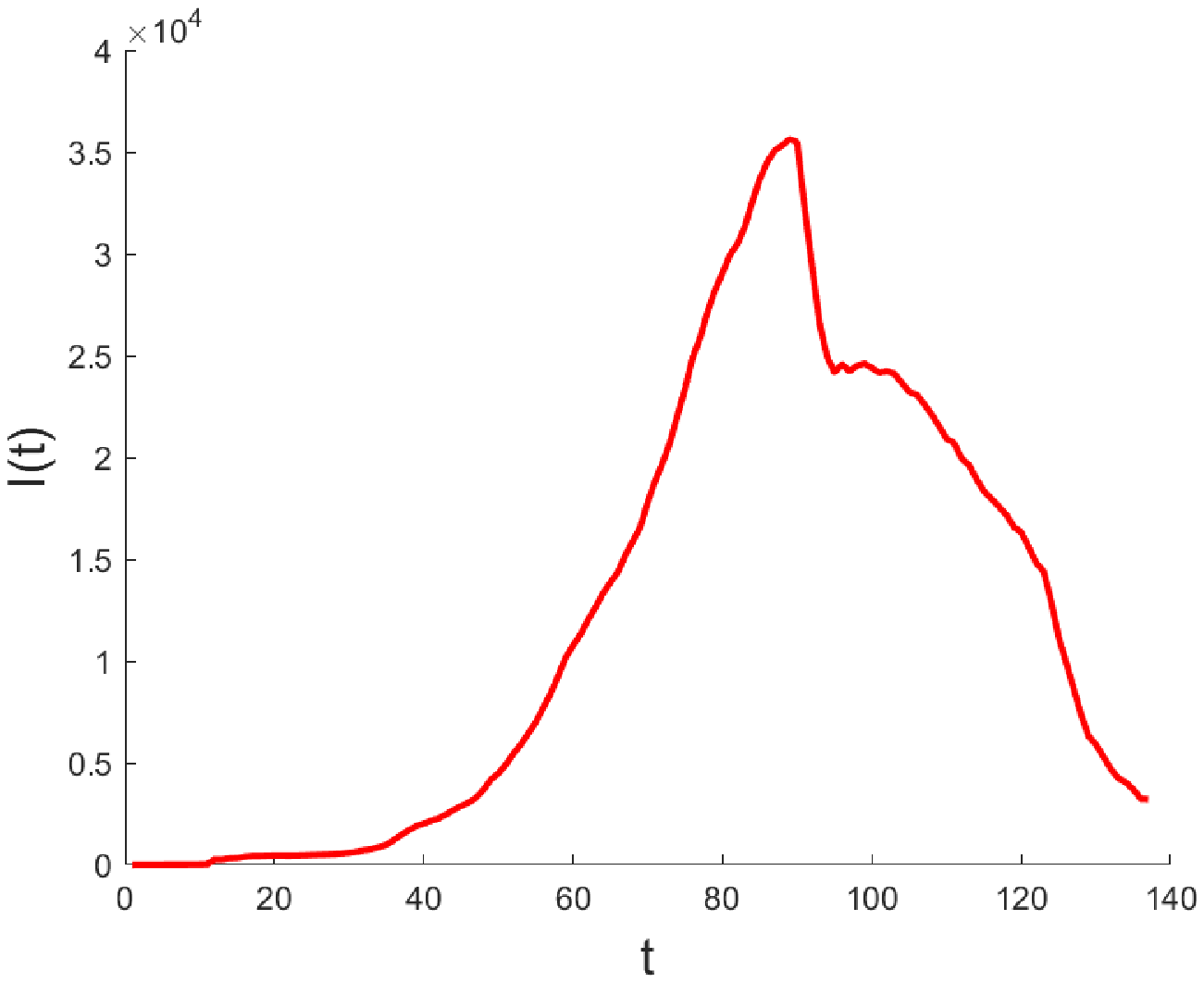}  &
\includegraphics[width = 0.31\textwidth]{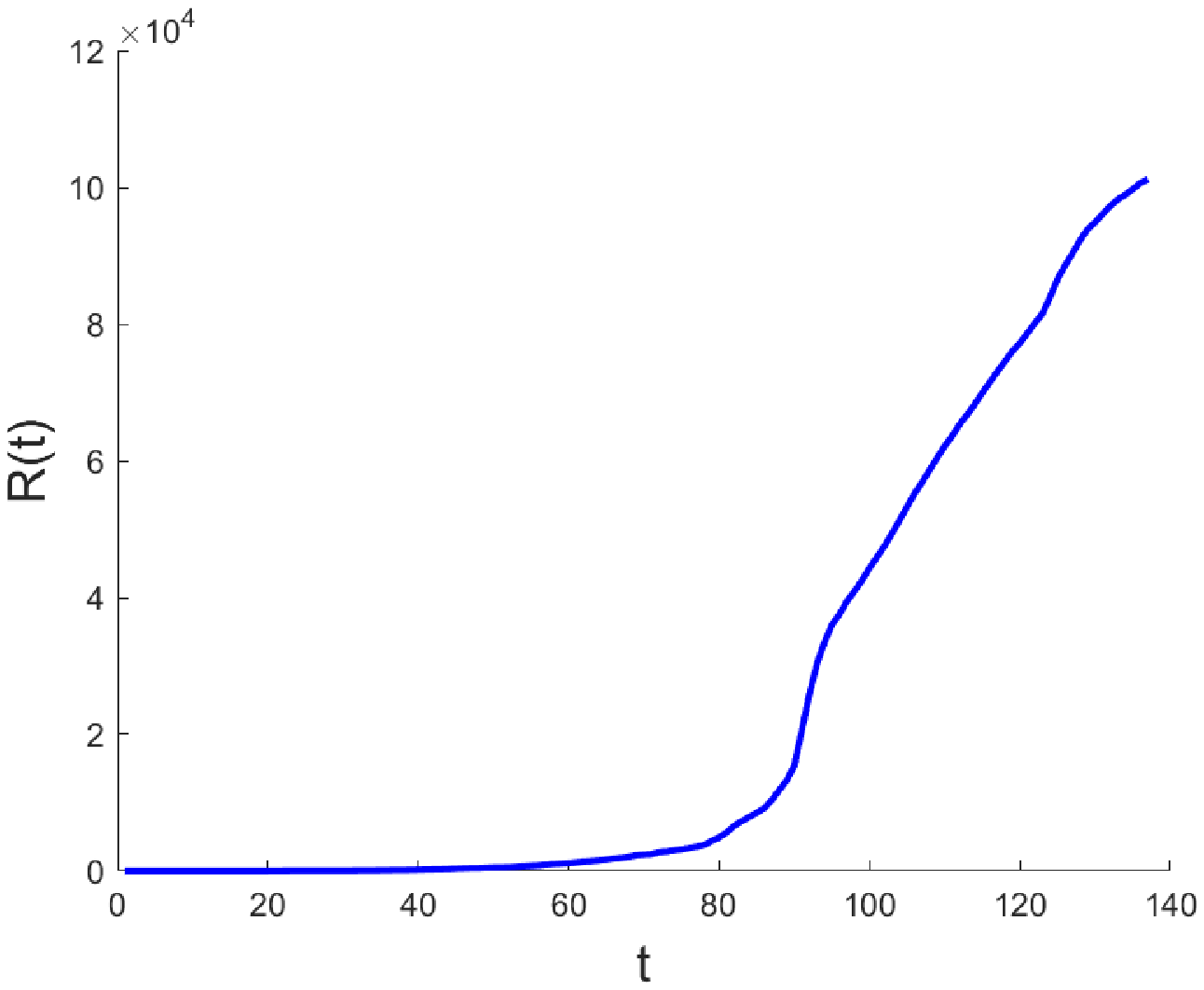}  &
\includegraphics[width = 0.31\textwidth]{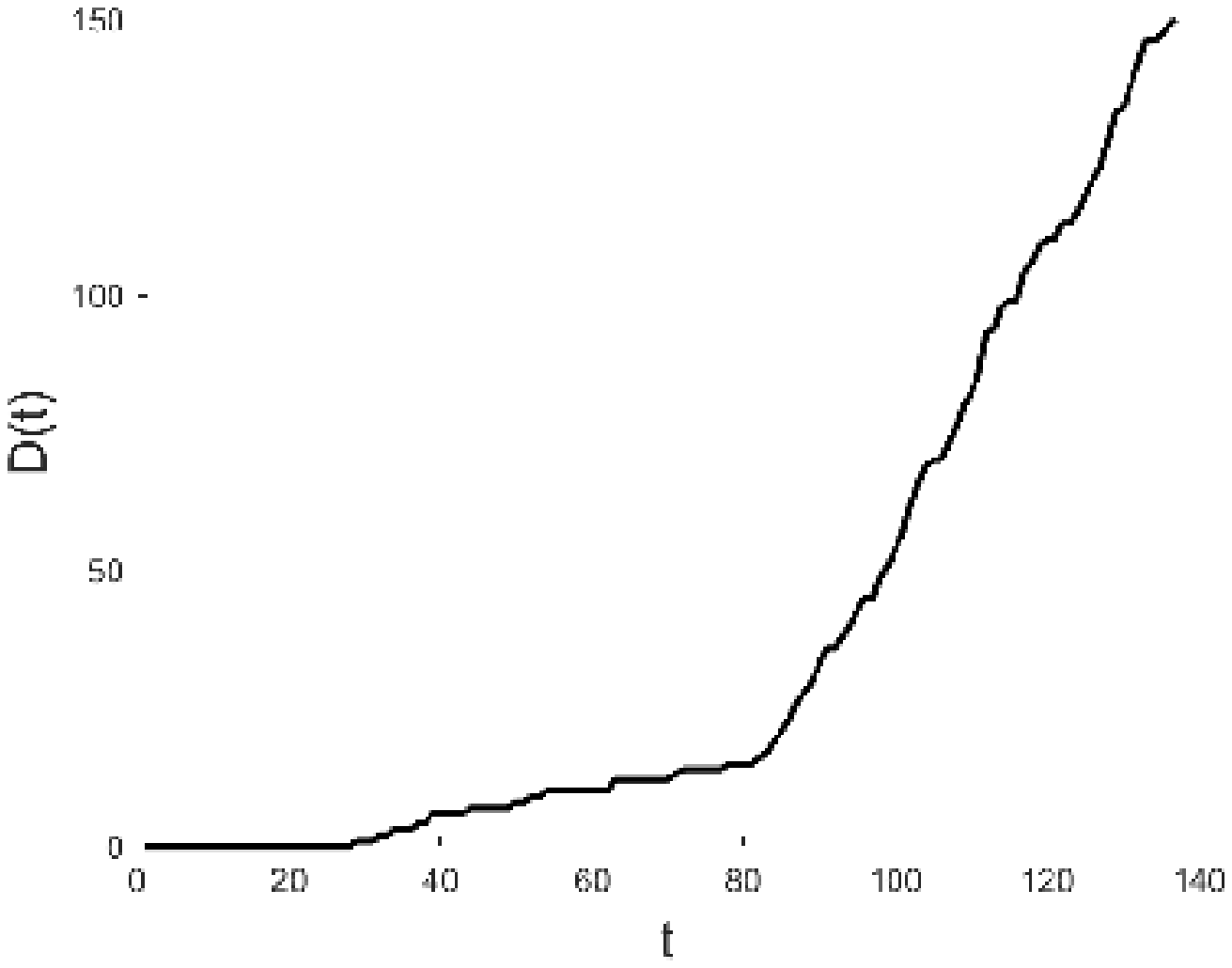}  \\
\end{tabular}
\caption{Plot of Active Infections (a), Recovered (b) and Deaths (c) for the State of Qatar data.} \label{fig:RawTQ}
\end{center}
\end{figure}

Figure \ref{fig:RawTQ} shows the plots of daily Active Infections, Recovered and Deaths data for the State of Qatar since 29 February 2020. The Active Infections start very low but encounter  a large jump around day 12 due to increased testing. The Active Infections then seem to plateau until day 30, after which there is an extreme growth in Active Infections. The Active Infections start to go down after day 90. The patterns in the Recovered and Deaths are similar and reflect the time of infection before recovery or death. Both graphs show a very slight increase in the number of recovered or dead subjects until about day 90, after which a steady increase is noticed.

\subsection{Evolution of the pandemic in Qatar}

The State of Qatar is one of the countries heavily affected by the COVID-19 pandemic. Since its first infected case way back on February 29, 2020, Qatar has become one of the highest infected countries in the Middle East with the total number of confirmed cases standing at 148,258 as of January 26, 2021. The total number of deaths in Qatar so far stands at 248 cases, which is low relative to the total number of infected cases, which is an indication of the country's highly effective healthcare system.

Qatar prepared an excellent flexible plan for risk management, grounded on national risk assessment, taking into account of the global risk assessment done by WHO and focusing on reinforcing capacities to reduce or eliminate health risks from COVID-19. Along with the well-organized healthcare system, the country was very quick to respond to the global pandemic. The country implemented many preventive measures very early on in the pandemic, including border control for early detection of cases. This included, were but not limited to, installing thermal screening for passengers who entered the country at Hamad International Airport and at seaports as early as January 2020, with the first quarantine facilities opening on February 1 \cite{Al Khal}.

On March 9, 2020 (day 10), Qatar closed all universities and schools and placed a travel ban on 15 countries: Bangladesh, China, Egypt, India, Iran, Iraq, Italy, Lebanon, Nepal, Pakistan, the Philippines, South Korea, Sri Lanka, Syria and Thailand. On March 14, 2020 (day 15), Qatar expanded its travel ban to include three new countries: Germany, Spain and France \cite{HMC, MPH}. The Ministry of Municipal and Environment on March 21, 2020, closed all parks and public beaches to curb the spread of coronavirus. On March 23, 2020 (day 24), the Ministry of Commerce and Industry decided to temporarily close all restaurants, cafes, food outlets, and food trucks for the public. Also, the Ministry of Commerce and Industry decided to close all unnecessary businesses on March 27, 2020 (day 28) \cite{HMC, MPH}.

As the number of infected cases continued to rise, on April 8, 2020 (day 40), the Ministry of Public Health (MoPH) announced that Primary Health Care Cooperation will be designating two health centers, one in Umm-Salal and one in Gharrafat Al-Rayyan, for screening, testing, and quarantining COVID-19 patients. MoPH also announced a hotline for psychological aid on April 9, 2020 (day 41). 

These interventions made by the government change the dynamics of the pandemic and hence, need to be considered while setting up a a real-time monitoring system of the infection, recovery and death rates. In the next subsection, we illustrate the proposed model as a data-driven forecasting model for use by stakeholders in the State of Qatar to monitor the COVID-19 pandemic.

\subsection{Data analysis results}

For Qatar, the prior distribution specification is:  $\alpha(t) \sim Exp( 2/4450000 )$, $\beta(t) \sim Exp( 1/105 )$, $\gamma(t) \sim Exp( 1/14 )$ and $\eta(t) \sim Exp( 1/9500 )$.  These priors reflect some information such as for $\gamma(t)$ the mean is $1/14$ which corresponds to a 14-day infection duration.

The SEIRD model was run with the following initial values for Qatar:  $S(0) = 2,782,000$, $E(0) = 3$, $I(0) = 1$, $R(0) = 0$ and $D(0) = 0$.  The SEIRD model and the sampling process  given in  Algorithm~\ref{alg:SIR} were coded in MATLAB R2020a and were run on a PC with Intel Core i7-7700 CPU at 3.60GHz with 8GB of RAM.  At each time step the sampler was run with $n_c = 10,000$, $n_p =1,000$ and $n_b = 10$.  Hence, each of the $n_p$ samples had a batch of $n_b$ samples generated in its neighborhood resulting in $n_p$ candidate particles at the beginning of the next time step.  The model computation time is about 60 minutes for $T=135$ time steps.  Note that, the number of individuals in each compartment in the model is much smaller due to the population size.  This means that many of the computations are faster especially when dealing with large factorials associated with the Poisson distribution.

\begin{figure}
\begin{center}
\begin{tabular}{cc}
(a) & (b) \\
  \includegraphics[width = 0.45\textwidth]{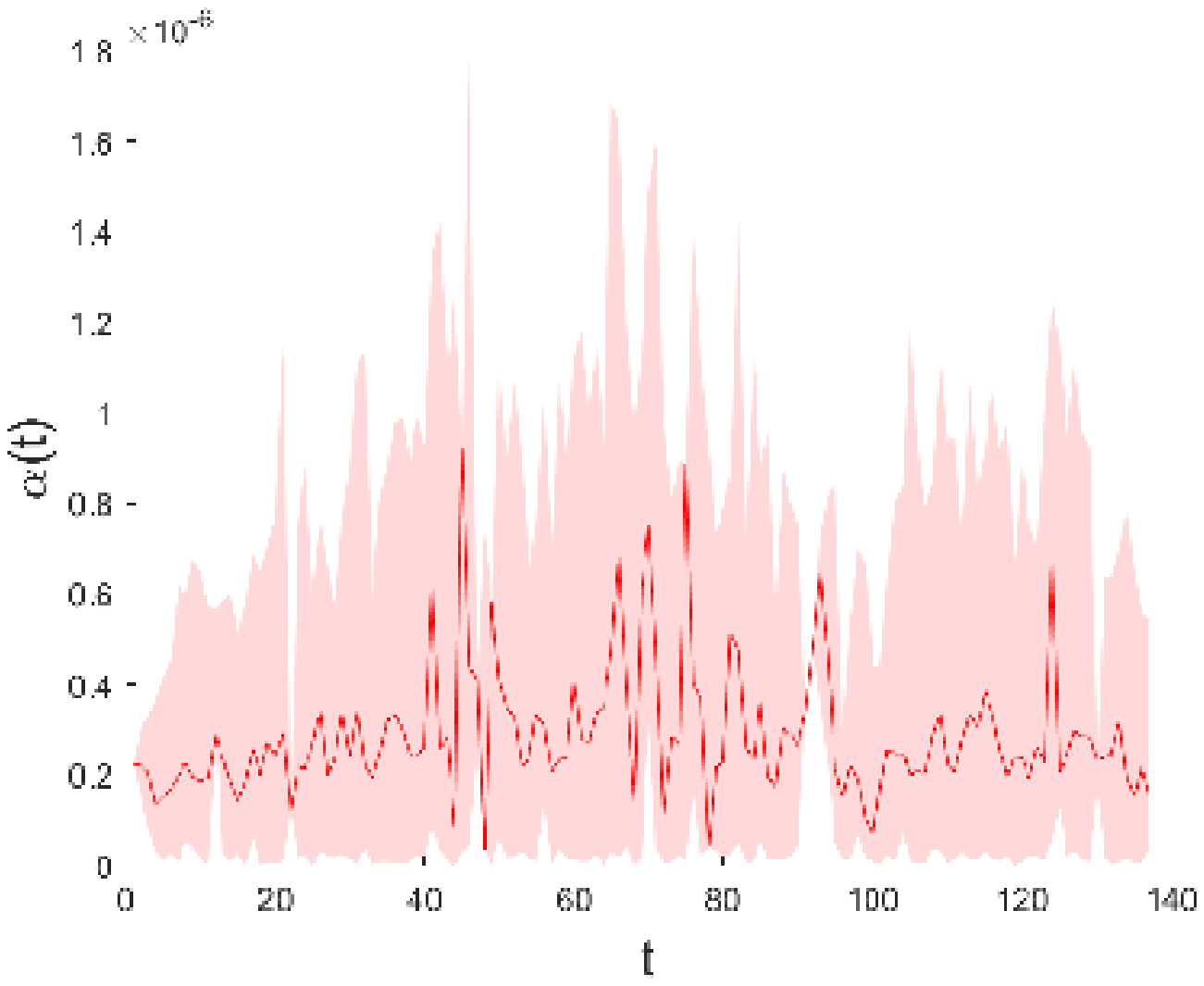}  &
  \includegraphics[width = 0.45\textwidth]{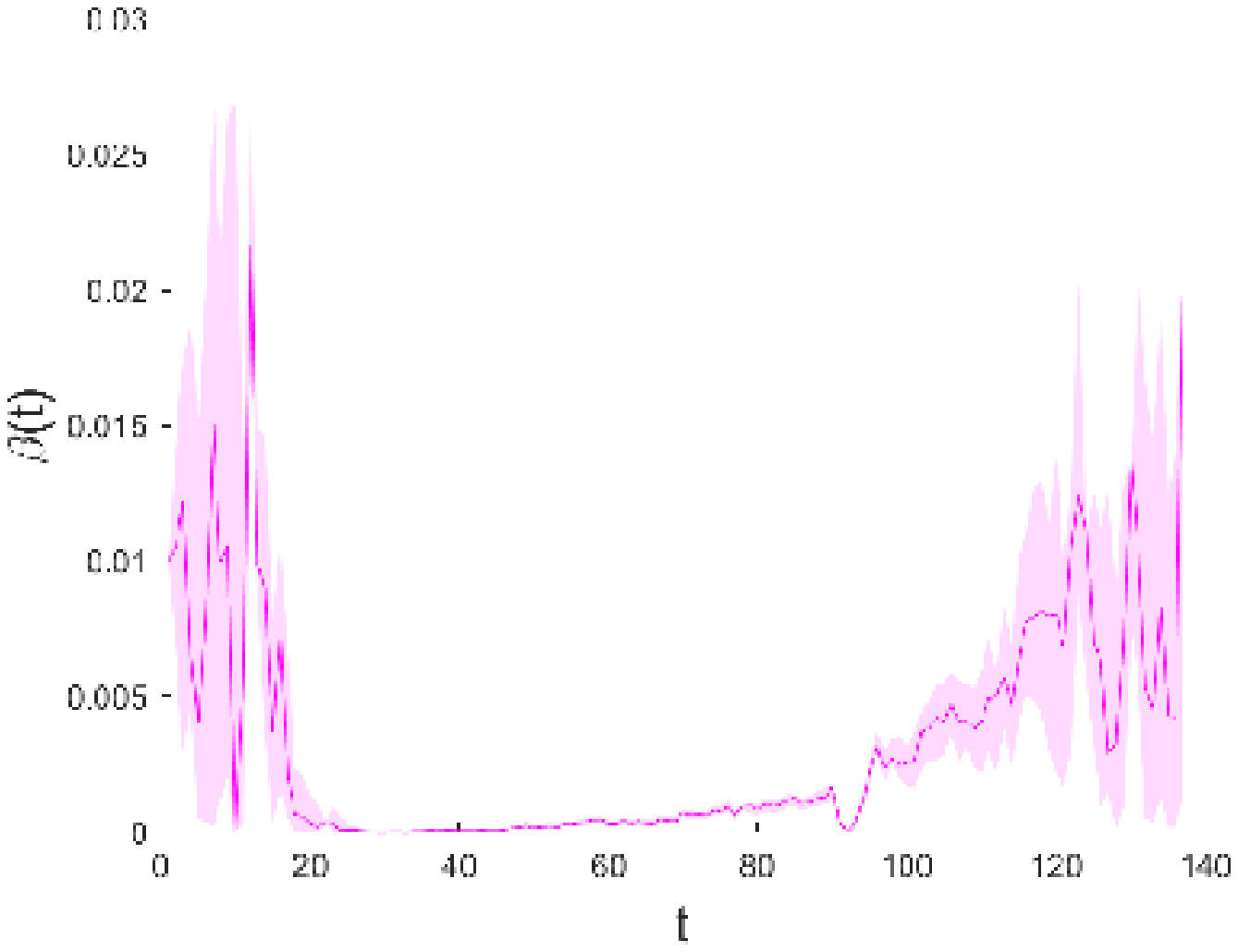} \\
  (c) & (d) \\
  \includegraphics[width = 0.45\textwidth]{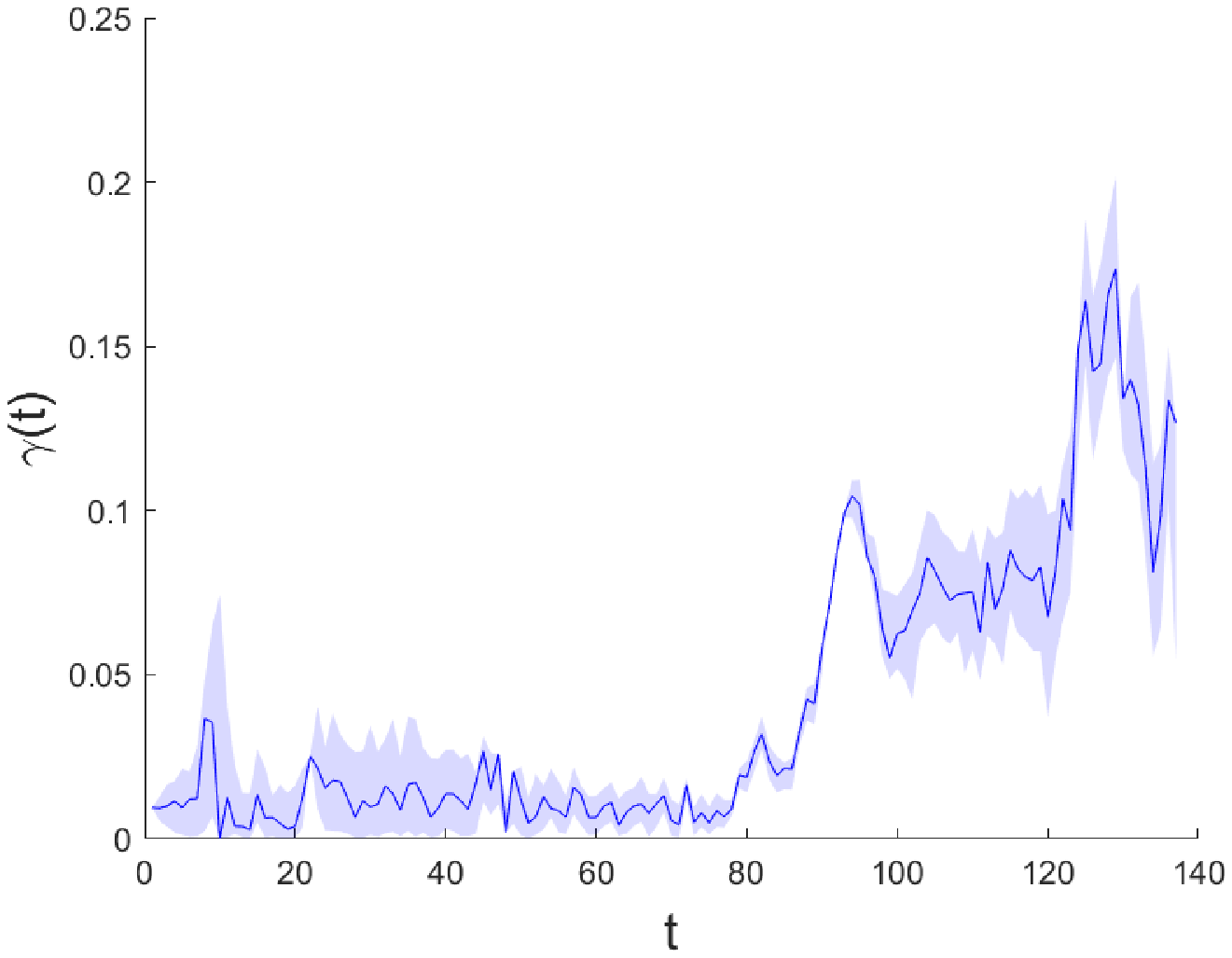} &
   \includegraphics[width = 0.45\textwidth]{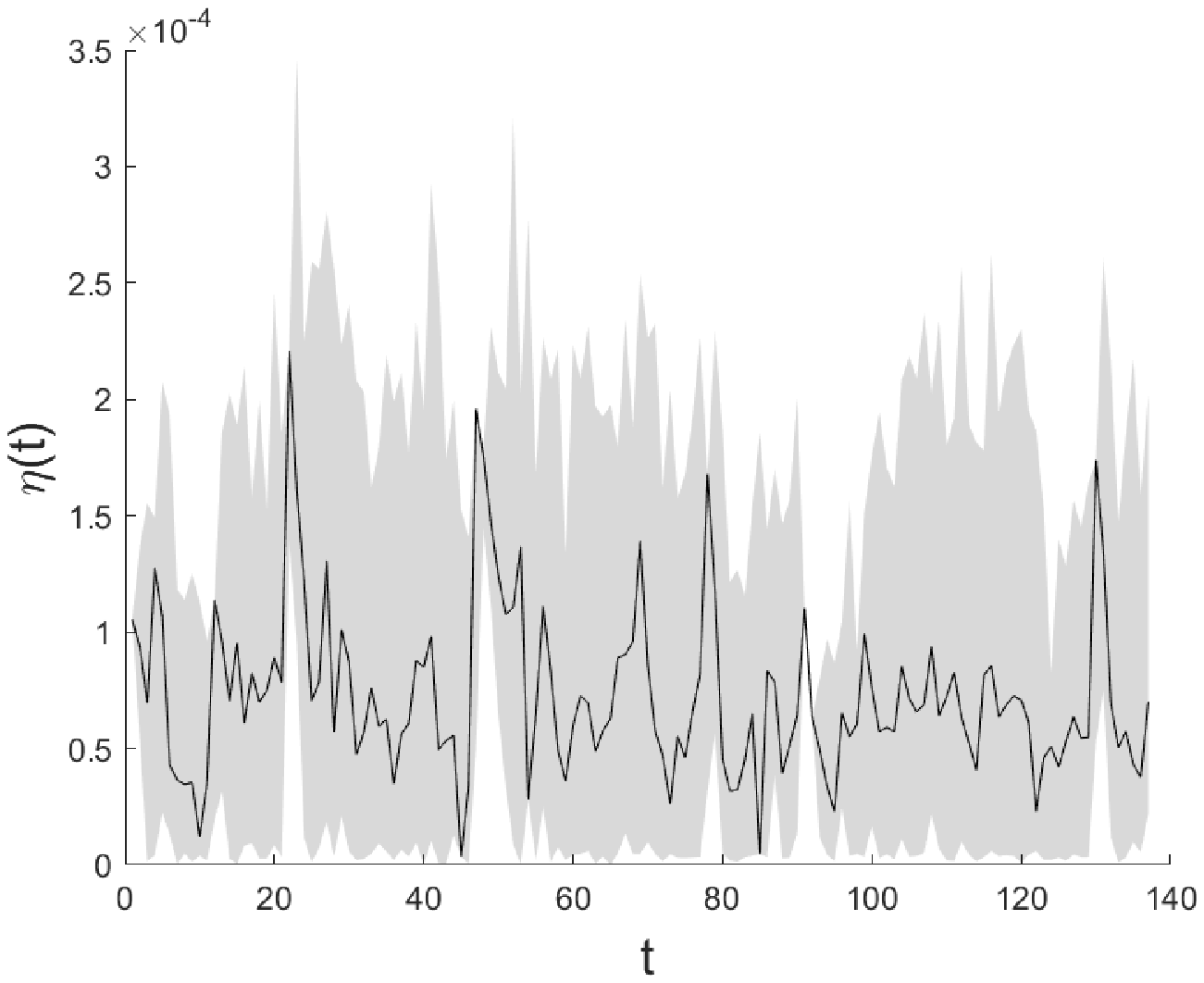}
 \end{tabular} 
\caption{Plots of $\alpha(t)$ (a), $\beta(t)$ (b), $\gamma(t)$ and $\eta(t)$ across time with associated 95\% credible intervals for the State of Qatar data.} \label{fig:QParTUS}
\end{center}
\end{figure}

 Figure~\ref{fig:QParTUS} shows the coefficient estimates across time, $\alpha(t)$ (panel a), $\beta(t)$ (panel b), $\gamma(t)$ (panel c) and $\eta(t)$ (panel d) for the Qatar data set.  Of particular interest is the time frame from day 90 to day 95 in which the active infections $I(t)$ exhibited a large drop (see Figure~\ref{fig:RawTQ}).  Notice that the distribution for $\alpha(t)$ becomes incredibly concentrated in this time frame as evidenced by the narrow credible intervals.  This is further exhibited in $\beta(t)$ and $\eta(t)$.  However, by examining $\gamma(t)$ during this time frame one sees that, a large spike in the recovery rate with very narrow credible intervals as well.  And for $\gamma(t)$ after about 80 days the recovery rate begins to increase dramatically with another spike around day 115.    

\begin{figure}
\begin{center}
\begin{tabular}{ccc}
(a) & (b) & (c) \\
\includegraphics[width = 0.31\textwidth]{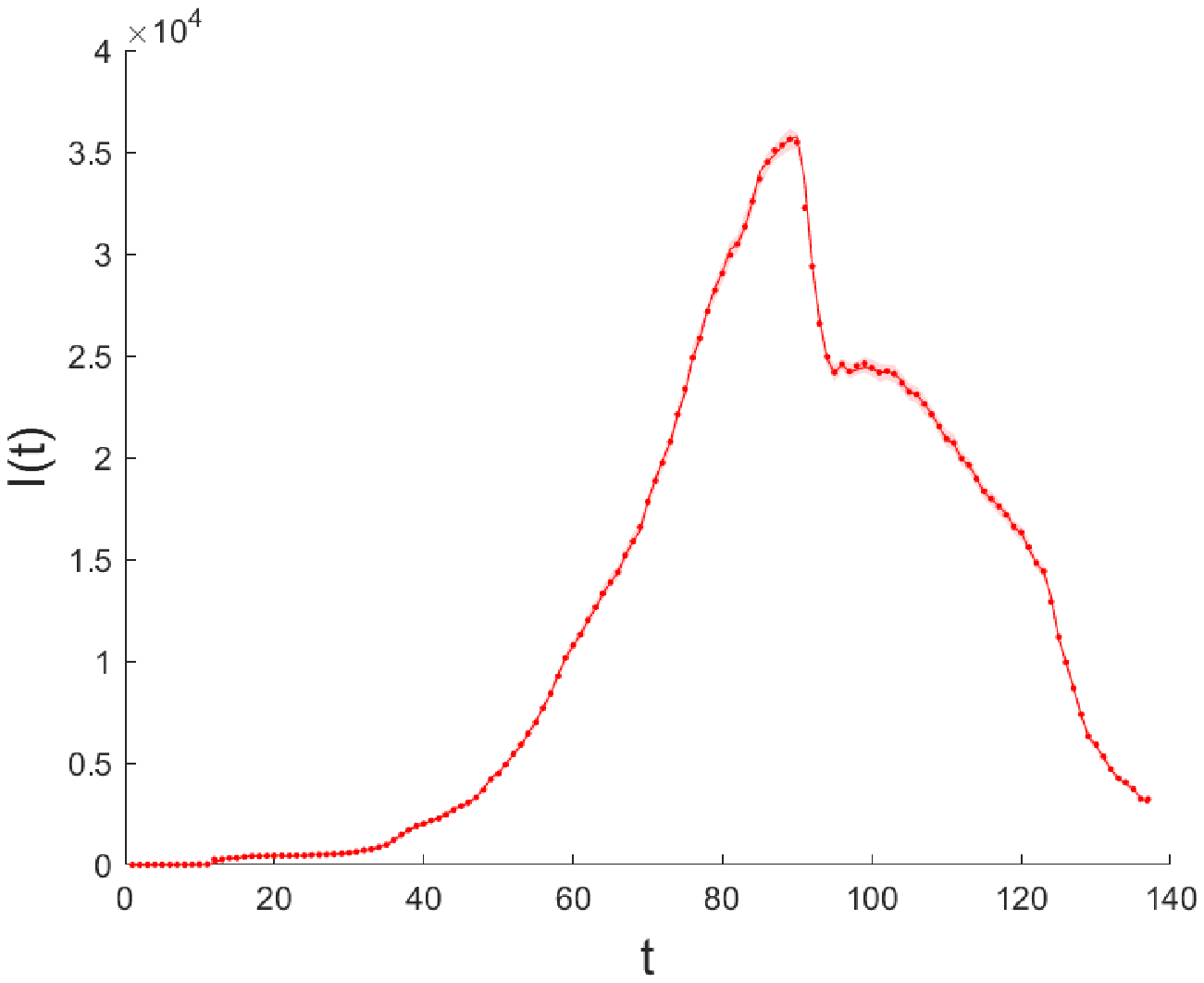}  &
\includegraphics[width = 0.31\textwidth]{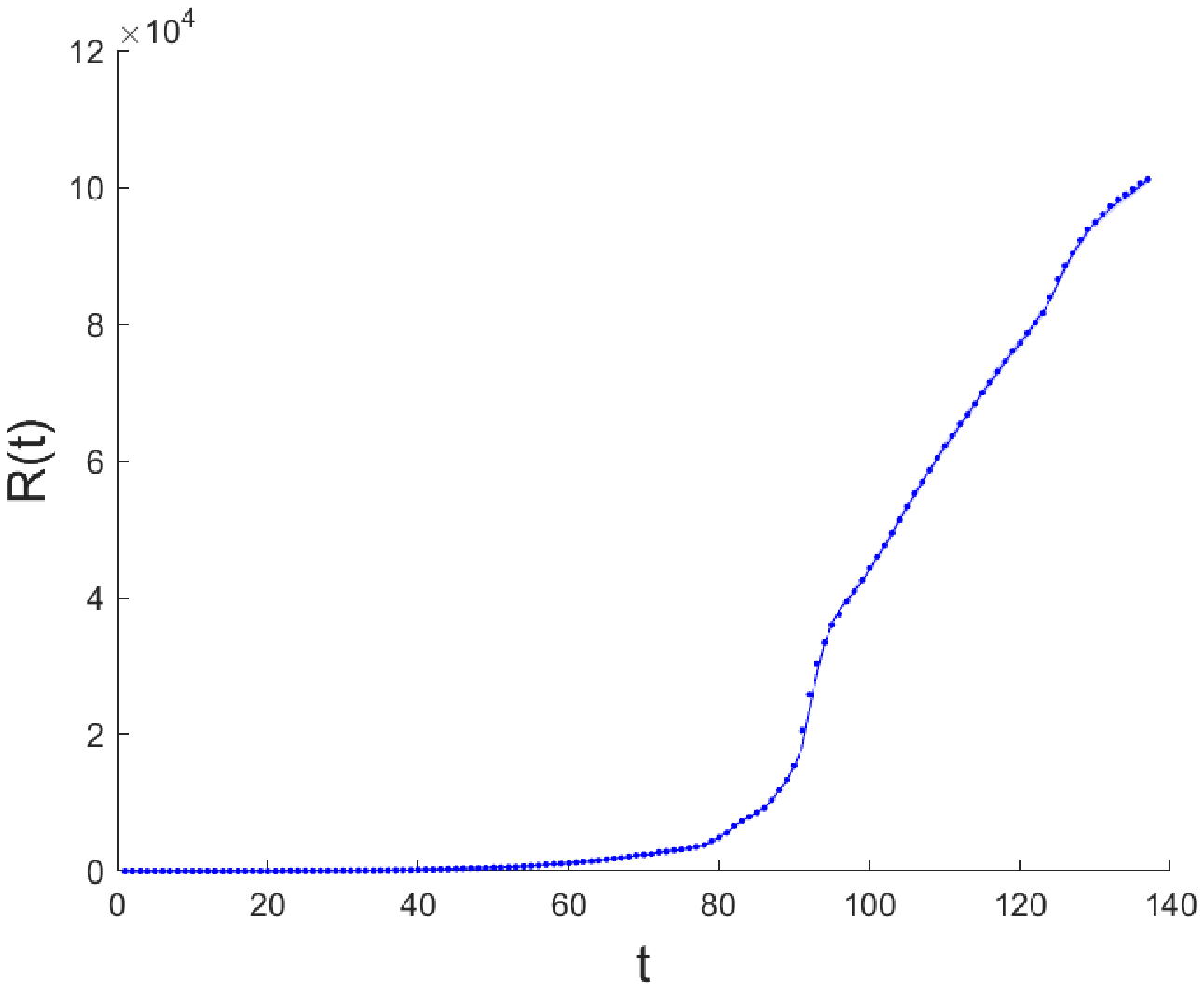}  &
\includegraphics[width = 0.31\textwidth]{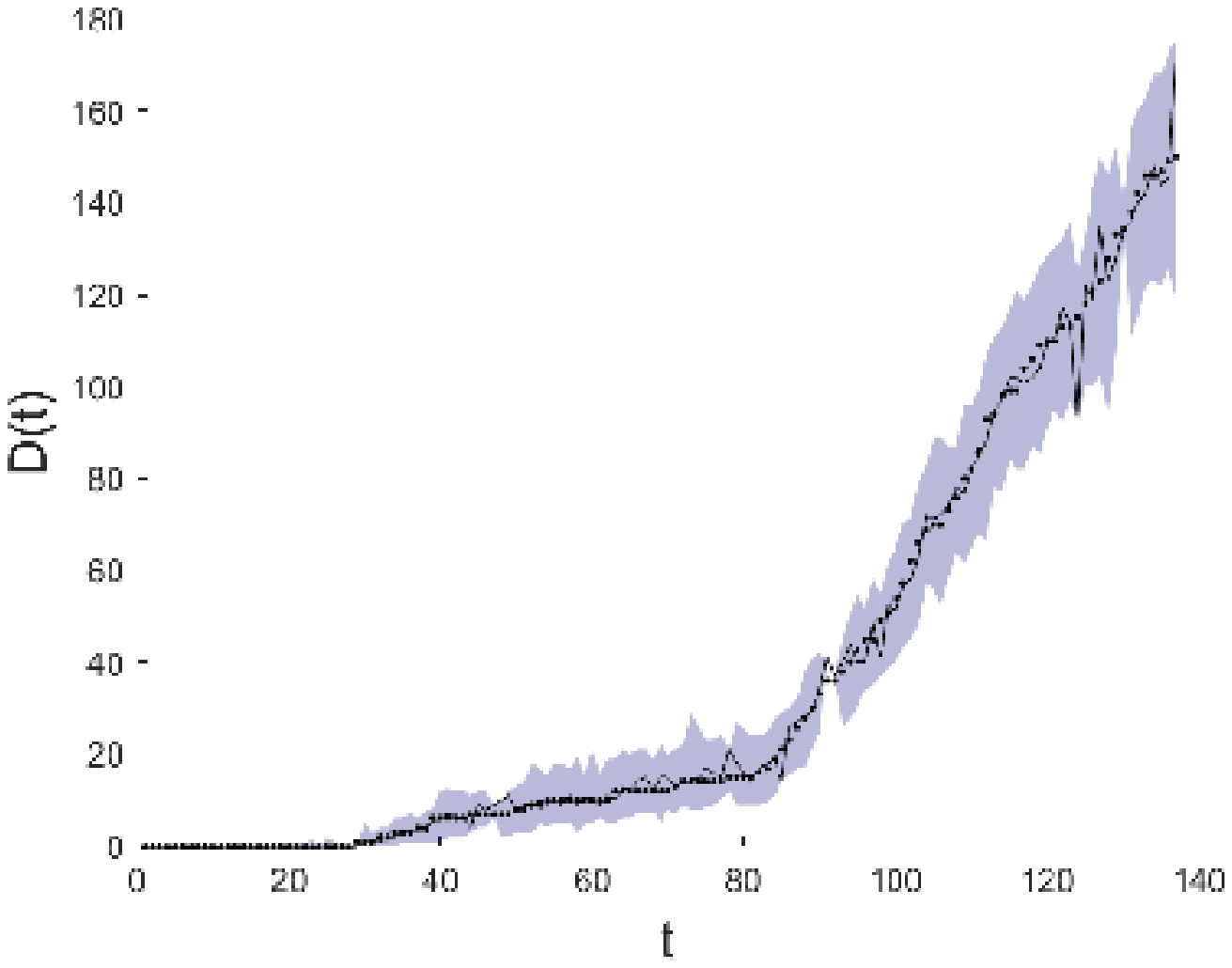}  \\
\end{tabular}
\caption{Plots of the data, fitted model with 95\% posterior prediction bounds for Active Infections (a), Recovered (b) and Deaths (c) for the State of Qatar data.} \label{fig:FitTQ}
\end{center}
\end{figure}

Figure~\ref{fig:FitTQ} shows the model fitted to the data with 95\% posterior predictive bounds for Active Infections (panel a), Recovered (panel b) and Deaths (panel c).  First note is that all three models appear to fit the data extremely well based on visual inspection.  Particularly notice that around day 90 when the dramatic drop in Active infections occurs this can be contrasted with Figure~\ref{fig:QParTUS}.

To assess the fit of the model a $Pseudo-R^2$ was calculated as:
\[ Pseudo-R^2 = 1- \frac{ \sum_{t=1}^{n} \left[ I(t) - \widehat{I}(t) \right]^2 + \sum_{t=1}^{n} \left[ R(t) - \widehat{R}(t) \right]^2 + \sum_{t=1}^{n} \left[ D(t) - \widehat{D}(t) \right]^2 }{\sum_{t=1}^{n} \left[ I(t) - \overline{I(t)} \right]^2 + \sum_{t=1}^{n} \left[ R(t) - \overline{R(t)} \right]^2 + \sum_{t=1}^{n} \left[ D(t) - \overline{D(t)} \right]^2  } \]
\noindent where $\overline{I(t)}$, $\overline{R(t)}$ $\overline{D(t)}$ are the sample means of $I$, $R$ and $D$, respectively, across time (and hence are not a function of time), $\widehat{I}(t)$, $\widehat{R}(t)$ and $\widehat{D}(t)$ are the medians of the posterior predictive distributions for $I$, $R$ and $D$, respectively, at each time ( and hence are functions of time),

Since uncertainty quantification is important, the proportion of observations that fall into the predictive bands was calculated as follows:
\[ \widehat{P}_{\text{fit}} = \frac{ \sum_{t=1}^{n} I_{\{I(t) \in \widehat{C}_{I(t)}\} } + \sum_{t=1}^{n} I_{\{R(t) \in \widehat{C}_{R(t)}\} }  + \sum_{t=1}^{n} I_{\{D(t) \in \widehat{C}_{D(t)}\} } }{ 3n } \]
\noindent where $\widehat{C}_{I(t)}$, $\widehat{C}_{R(t)}$ and $\widehat{C}_{E(t)}$ denote the 95\% predictive intervals for $I(t)$, $R(t)$ and $D(t)$, respectively and  $I_{\lbrace \mathcal{A} \rbrace} $ is an indicator function taking the value one if event $\mathcal{A}$ is true.

The \emph{Pseudo-R$^2 = 0.9999$} which shows an incredible agreement between the  median fitted values and their corresponding data values.  The proportion of observations that fall into the 95\% predictive bands $\widehat{P}_{\text{fit}} = 0.8394$ which indicates that the model has less uncertainty that it should.  However, the value is still quite high with approximately 84\% of observations being captured by the intervals.

\begin{figure}
\begin{center}
\includegraphics[width = 0.66\textwidth]{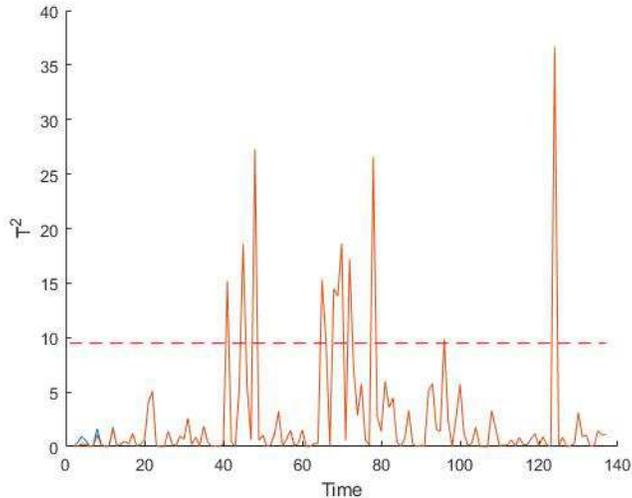}  
\caption{Plot of the Hotelling's $T^2$ statistic through time.  Horizontal line corresponds to the 95\% control limits} \label{fig:MEWMA1}
\end{center}
\end{figure}

Figure~\ref{fig:MEWMA1} shows the plot of $T^2(t)$ for the Qatar data with a control limit set at 9.48 (red dashed line) and a smoothing parameter $\Lambda = 0.2$.  Notice that until time 40 the process seems to be pretty stable as indicated by the $T^2(t)$ being below the control limit.  After day 40 there are several time points signalling a change in the process on days: 40, 44, 47, 64, 65, 67, 68, 69, 71, 77, 95, and 123.  The early days (40,44,47) can easily be seen to agree with the change in both infection rate $\alpha(t)$ and death rate $\eta(t)$ in Figure~\ref{fig:QParTUS}  (a) and (d).  Notice in these plots the high volatility with spikes in $\alpha(t)$ and a clear shift upwards in $\eta(t)$ at the same times.  During the days 64 to 77 there appears to be a very large amount of volatility in the infection rate $\alpha(t)$ which is also evidenced in Figure~\ref{fig:FitTQ} (a) for active infections.  Notice how the active infections have large increases one day and smaller increases the next day.  The method also picks up the spike in infection rates as well as the dramatic increase in recovery rate $\gamma(t)$ at time 95.  At day 123, Figure~\ref{fig:QParTUS} shows a spike in the rate of exposed to actively infected, $\beta(t)$ (panel b) and another shift in recovery rate $\gamma(t)$ (panel c). 

To study the sensitivity to the smoothing parameter in the MEWMA chart across signals several other analyses were performed with $\Lambda = 0.1, \Lambda = 0.15, \Lambda = 0.25$, and $\Lambda = 0.3$, which are commonly chosen values for this parameter.  When $\Lambda = 0.1$ the monitoring process signaled at days: 44, 47, 69, 77 and 123.  Since the smoothing parameter puts more weight on the previous mean than on the new observations, it is expected that a smaller number of days would be signalled.  When $\Lambda = 0.15$ the monitoring process signaled at days: 40, 44, 47, 64, 68, 69, 71, 77 and 123.  Notice that for both $\Lambda = 0.1$ and $\Lambda = 0.15$ day 95 is not signaled, which upon inspection of almost all the charts there is a shift in the process.  Hence, these parameter choices would be considered too conservative in this case.  When $\Lambda = 0.25$ the following days were signalled:  40, 44, 47, 64, 65, 67, 68, 69, 71, 77, 95 and 123.  This is the same result as when $\Lambda = 0.2$.  When $\Lambda = 0.3$ the monitoring process signalled the following days: 40, 44, 47, 64, 65, 67, 68, 69, 71, 72, 74, 77, 80, 95 and 123.  Here days 72, 74 and 80 are added to the list of signalled days.  This reflects the volatility in $\alpha(t)$ across this time frame.  Overall, when less conservative values for $\Lambda$ are chosen, the days signalled are quite reasonable.  In could be argued that in a pandemic situation a more sensitive monitoring process would be beneficial to public policy makers as it can signal when an effective intervention has been introduced.  

%
%
\section{Discussion}\label{sec:Discuss}

This work provides a novel tool for monitoring and capturing changes in a pandemic evolution process via monitoring changes in parameters of mathematical epidemiological models, such as the Susceptible, Exposed, Infected, Recovered, Death (SEIRD) model using the Multivariate Exponentially Weighted Moving Average (MEWMA) process monitoring technique.  A Bayesian approach is taken for the parameter estimation with a sampling algorithm that allows for both quick updating of the SEIRD model but also provides samples that can be monitored by the MEWMA regime. This sampling algorithm uses the notion of Sampling Importance Resampling, but augments the particles at each step to avoid particle depletion. This quick updating allows for the process monitoring scheme to “signal” quickly if there is a change in the model parameters. The method is then used to monitor the evolution of the COVID-19 pandemic in the State of Qatar. 

Despite the proliferation of forecasting models for the evolution of the COVID-19 pandemic, their accuracy achieved  can be compromised and  comparisons can be complicated due to numerous factors, e.g., their construction methods, distinct healthcare systems adopted by different countries/regions, different political decisions or policies made, distinct testing and reporting mechanisms \cite{Nikolopoulos2021}. Hence, using the forecasts given by a particular forecasting model for critical decision making is challenging. 
The proposed approach takes a different perspective and enables decision-makers to work with a tailored SEIRD model, assess the effectiveness of the policies/decisions made, and adopt interventions and/or prevention strategies consistently over time. 

The State of Qatar example illustrates the proposed method's ability to perform daily monitoring of a pandemic. The proposed model fits the data very well with a $pseudo-R^2 =  0.999$. In the model definition, immigration, emigration, natural births and natural mortality have not been included; however, based on the high psuedo-$R^2$, they would have a negligible effect on the fit. Furthermore, the model does not contain compartments for subjects who recovered without being confirmed infections. Since this is not observed, one can only speculate on the impact that additional data would have on the model fit; however, it would be very small. As seen in Figures~\ref{fig:QParTUS} and \ref{fig:FitTQ}, the proposed method successfully picks up the day to day fluctuations in the pandemic evolution process in Qatar via the estimated time-varying model parameters. Note that the pandemic's overall state can also be monitored by tracking the $T^2$ statistic over time (see Figure~\ref{fig:MEWMA1}). For Qatar, the method signals the first change in the process around day 40. This change can be attributed to several government interventions such as closing parks and public beaches on day 24, closing all unnecessary businesses on day 28 and announcing two major health centers catered towards COVID-19 patients on day 40. The method also signals multiple days beyond day 40, all of which seem reasonable upon further inspection. Thus, the proposed method gives decision-makers the ability to evaluate planned interventions as well as discover new changes to the process and respond accordingly. This method can also be extended for monitoring a process at the state/county level by incorporating a spatial covariance and using the mixed model approach. 

Since the augmented sampling regime allows posterior samples to be saved from the previous day, updating is performed on a daily basis and only requires the new data and the previous day's samples.  Thus the entire SEIRD model need not be fit from the beginning of the series.  Furthermore, the MEWMA is quickly calculated from the posterior samples and can quickly signal those managing the pandemic. Note that the method is not tied to the SEIRD model given in Equation (\ref{eq:Sys1}), as the augmented sampler and MEWMA monitoring protocol are generic.  Our motivation here has been using a system where the reproduction number fails to include all the relevant parameters.  In systems where the reproduction number is dependent on all parameters, the reproduction number could be added as a dimension to the monitoring protocol as well.  In situations where the reproduction number is meaningful, this could be another dimension that could ``signal'' serious changes in long-term process outcomes.



\end{document}